\documentclass[fleqn,12pt,a4paper]{article}
\usepackage{espcrc1}
\usepackage{graphicx,epsfig}
\usepackage{amsmath}

\usepackage{makeidx}  
\usepackage{times}
\usepackage{amsfonts}
\usepackage{amssymb}
\usepackage{amsmath}
\usepackage{xspace}
\usepackage{subfigure}

\title{Numerical Approach to Calculation of Feynman Loop Integrals\footnote{Talk given at 3rd Computational Particle Physics Workshop – CPP2010, September 23-25, 2010, KEK
Japan}}

\author{F. Yuasa\footnote{E-mail: fukuko.yuasa@kek.jp}\address[KEK]{High Energy Accelerator research Organization (KEK), 1-1 Oho Tsukuba, Ibaraki 305-0801, Japan}, T. Ishikawa\addressmark[KEK], Y. Kurihara\addressmark[KEK], J. Fujimoto\addressmark[KEK], Y. Shimizu\addressmark[KEK], N. Hamaguchi\addressmark[KEK], \\E. de Doncker\address{Western Michigan University Kalamazoo, MI 49008-5371, USA} and K. Kato\address{Kogakuin University, 1-24 Nishi-Shinjuku, Shinjuku, Tokyo 163-8677, Japan}
}

\begin{document}

\maketitle       

\begin{abstract}
In this paper, we describe a numerical approach to evaluate 
Feynman loop integrals. 
In this approach the key technique is a combination of a numerical integration method 
and a numerical extrapolation method. Since the computation is carried 
out in a fully numerical way, our approach is applicable to 
one-, two- and multi-loop diagrams. Without any analytic treatment it can compute 
diagrams with not only real masses but also complex masses for the internal 
particles. As concrete examples we present numerical results of 
a scalar one-loop box integral with complex masses and two-loop planar and non-planar 
box integrals with masses. We discuss the quality of our numerical 
computation by comparisons with other methods and also propose a self consistency check.
\end{abstract}


\section{Introduction}\label{sec:intro}

\noindent
A method for the systematic calculation of loop diagrams is required to get 
precise theoretical predictions for elementary particle interactions. 
In this paper we present a fully numerical approach toward this computation. 
For simplicity we consider scalar loop integrals throughout the paper.
The general expression of scalar loop integrals in Feynman parametric representation is
\begin{equation}\label{eqn:general}
(-1)^{N}\frac{\Gamma(N-nL/2)}{(4\pi)^{nL/2}}
\int_{0}^{1} \prod_{i=1}^{N} \,dx_{i}~\delta(1-\sum_{i=1}^{N}x_i)\,\frac{C^{N-n(L+1)/2}}{(D-i\epsilon C)^{N-nL/2}}
\end{equation}
where $L$ is the number of loops, $N$ is the number of internal particles and $n$ is the
number of space-time dimensions. Here, $C$ and $D$ are polynomials of the 
Feynman parameters ($x_i, ~i=1, \cdots, N$) and they are determined by the topology of the corresponding diagram.
An infinitesimal parameter, $\epsilon$, is introduced to make the denominator non-zero throughout the integration domain.
With $n=4$, the form of the expressions of the integrand for 
each $N$ and $L$ is shown in Table~\ref{tab:integrand}.

\begin{table}[h]
\label{tab:integrand}
\caption{Integrand of scalar loop integrals for the case $n=4$ up to $L=2$ and $N=7$. For $L=1$, $C\equiv 1$.}
\begin{center}
\begin{tabular}{llc}\hline
L& N & $C^{N-2(L+1)}/(D-i\epsilon C)^{N-2L}$ \\ \hline
1 & 3 &  $1/(D-i\epsilon)$\\
  & 4 &  $1/(D-i\epsilon)^{2}$\\
  & 5 &  $1/(D-i\epsilon)^{3}$\\ \hline
2 & 5 &  $1/(C(D-i\epsilon C))$\\
  & 6 &  $1/(D-i\epsilon C)^{2}$\\
  & 7 &  $C/(D-i\epsilon C)^{3}$\\ \hline
\end{tabular}
\end{center}
\end{table}

A standard (analytic) method for multi-loop integrals is by a reduction to a set of integrals using integration 
by parts~\cite{tkachov}. However, reduction often yields a large number of loop 
integrals and it becomes difficult to obtain an accurate numerical
result due to stability problems and cancellation error, with a loss of trailing digits. 
On the other hand, in our approach it is not necessary to reduce the integral and the number of integrals to be computed is limited.

Generally speaking the numerical computation of loop integrals becomes harder 
with an increasing number of loops and/or external legs
since the behavior of the singularities becomes more complicated. 
We use an automatic integration technique with an efficient  
numerical extrapolation method and, if necessary, 
a suitable variable transformation of the Feynman parameters.
Since all computation is completely numerical, it is trivial in our approach to extend 
the number of loops and legs, be it at the price of an increased amount of work
as measured in the number of integrand evaluations.
Another advantage of the general numerical approach is that it does not matter whether 
the masses of the internal particles are real or complex.
So far we have applied our approach to the computation of one-loop vertex ($L=1, ~N=3$), 
box ($L=1, ~N=4$) and pentagon ($L=1, ~N=5$) as well as two-loop selfenergy ($L=2, ~N=5$)
and vertex ($L=2, ~N=6$) diagrams~\cite{dq1,dq2,dq3,dq4,dq5,acat08}.
Numerical results by our method show good agreement with that of other methods~\cite{1loop1,supplement,1loop2,1loop3,kreimer,2loopself,2loopv1,2loopv2,tarasov,bauberger,passarino-self,passarino-2lv,nci,ueda-acat08}.

Subsequently in section~\ref{sec:directcomputation} we give a brief explanation of our 
approach with respect to the numerical techniques.
In section~\ref{sec:results} we present results of one-loop and two-loop diagrams with 
real or complex masses up to $L=2, ~N=7$ as examples.  

\section{Direct Computation Method}\label{sec:directcomputation}

\noindent
As in Eq.\,(\ref{eqn:general}), $i\epsilon$ is introduced in the denominator of the integrand.
For instance, for the case $L=1$ and $N=3$ in Table~\ref{tab:integrand}, we separate 
the real and imaginary part of the integrand as
\begin{align}
\Re e \frac{1}{D - i\epsilon}&=\frac{D}{D^2+\epsilon^2} \label{eqn:re}\\
\Im m \frac{1}{D - i\epsilon}&=\frac{\epsilon}{D^2+\epsilon^2} \label{eqn:im}
\end{align}
since our numerical integration package {\tt Quadpack}~\cite{quadpack} 
currently does not support complex integrands.
While analytically $\epsilon$ is thought of as infinitesimal, 
we will replace it by a sizable number in a sequence of the form $\epsilon = \epsilon_j = a^{(l-j)}, j=0, 1, \cdots$
where $a$ is a positive number and $l$ is an integer.
When $\epsilon_j$ is finite, the integral converges and numerical methods can be applied to the integration of Eq.\,(\ref{eqn:re}-\ref{eqn:im}).
For instance, varying $\epsilon_j$ geometrically as $1.2^{(30-j)}$ with $a=1.2$ and $l=30$, we get a sequence of integrals $I(\epsilon_j)$
corresponding to each $\epsilon_j.$
It is our goal to obtain the limit $\lim_{\epsilon_j \rightarrow 0} I(\epsilon_j).$ 
This is an extrapolation as $\epsilon_j \rightarrow 0$ and we can accelerate the convergence of the sequence 
by an appropriate acceleration technique.
We refer to this method, {\it i.e.}, the combination of the integration and the extrapolation, as a {\it Direct Computation Method}.

\subsection{Multi-dimensional integration}\label{subsec:integration}

\noindent
Basically in our method, we can choose any numerical integration procedure
if it gives the numerical result to enough accuracy.
However, most numerical integration methods fail due to singularities in the integration domain.
We use the {\tt DQAGE} routine from the one-dimensional {\tt Quadpack} package~\cite{quadpack} 
for a repeated integration~\cite{dq1,dq2,iterate} in the coordinate directions of the
multi-dimensional integral.
{\tt DQAGE} is an adaptive quadrature routine. Generally it will partition around an integrand 
singularity ``hot-spot" within the integration region, for an arbitrary location of the singularity.
On each subinterval generated in the subdivision, {\tt DQAGE} applies a variant of Gaussian quadrature where the
sampling points are chosen by a Gauss-Kronrod scheme. 
  
\subsection{Extrapolation}\label{subsec:extrapolation}
 
\noindent
As for the extrapolation, we choose an extrapolation method which does not require explicit 
information to be supplied about the sequence.
Throughout this paper we present results using Wynn's $\varepsilon$ algorithm~\cite{wynn} as 
the extrapolation method.
This works very efficiently even for sequences of $I(\epsilon_j)$ with slow convergence 
(providing the progression of $\epsilon_j$ is geometric).

\section{Computation of the loop integrals}\label{sec:results}

\noindent
Here we show some numerical results by the {\it Direct Computation Method}.
Since our approach is based on the complete numerical technique, the computation is possible for loop integrals with arbitrary masses no matter whether they are real or complex.
The first example, in section~\ref{subsec:1lb}, is a scalar one-loop box diagram with complex masses. 
In section~\ref{subsec:2lb}, we treat scalar two-loop planar and non-planar box diagrams with real masses.
For these examples we not only present numerical results but also discuss a technique 
for a cross-check of our numerical results.
There are several ways for cross-checking, and comparisons with results by other methods are effective.
However, if no other results are available for comparison, we propose a self consistency check 
to test the quality of the computation.

\subsection{One-loop box with complex masses}\label{subsec:1lb}

\begin{figure}[h]
\label{fig:box}
\begin{center}
\includegraphics[width=6.0cm]{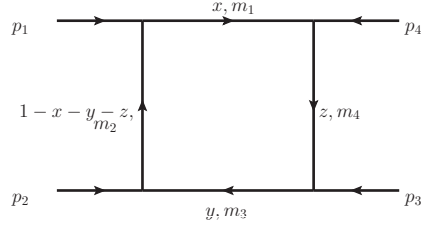}
\end{center}
\caption{One-loop box diagram}
\end{figure}

\noindent
In this section we consider a scalar integral $I(s,t)$ for the one-loop box diagram shown in Fig.~\ref{fig:box} ($L=1$, $N=4$ in Table~\ref{tab:integrand}) defined as 

\begin{equation}\label{eqn:integral-1lb}
I(s,t) = \int_{0}^{1}dx\int_{0}^{1-x}dy\int_{0}^{1-x-y}dz \frac{1}{(D-i\epsilon)^2},
\end{equation}
where $D$ is given by
\begin{eqnarray*}
D &=& p_1^2x^2+p_2^2y^2+tz^2 + (p_1^2+p_2^2-s)xy + (p_1^2-p_4^2+t)xz+(p_2^2-p_3^2+t)yz \\
  &+&(-p_1^2+m_1^2-m_2^2)x+(-p_2^2+m_3^2-m_2^2)y+(m_4^2-m_2^2-t)z +m_2^2,
\end{eqnarray*}
with $s=(p_1+p_2)^2=(p_3+p_4)^2$ and $t=(p_1+p_4)^2=(p_2+p_3)^2$.
One-loop box integrals have been fully analyzed and several powerful tools such as {\tt FF}~\cite{FF} and {\tt LoopTools}~\cite{looptools} have been
developed for their numerical evaluation. However, it is often tedious to include
complex masses for the internal particles of the box diagram analytically.

In 2007 and 2008 we compared our results for the one-loop box diagram contribution of  
$gg \rightarrow b{\bar b}H,$ to those obtained analytically by L.\,D. Ninh 
{\it et al.}~\cite{bbh}.
A severe numerical instability due to a Landau singularity was reported by L.\,D. Ninh for $211 \,\rm {GeV} \geq M_H \geq 2M_W$ and $457\,\rm{GeV} \geq \sqrt s \geq m_t$ in his numerical evaluation; and we observed the same instability using the {\it Direct Computation Method}.
To regularize the singularity we, as well as L.\,D. Ninh and co-authors, included 
two complex masses for the internal particles as $m_t^2 -i m_t \Gamma_t$ and 
$M_W^2 - i M_W \Gamma_W,$ with $\Gamma_t =1.5$ GeV and $\Gamma_W=2.1$ GeV, respectively. 
After inclusion of the widths the instability disappeared and 
the results by L.\,D. Ninh showed very good agreement with ours~\cite{acat08}. 

Subsequently in 2010 we computed the one-loop box integral with complex masses set to
$m_1^2=20-0i$, $m_2^2=10-5i$, $m_3^2=40-10i$ and  $m_4^2=10,100,1000$ with $p_1^2=-60$, $p_2^2=10$, $p_3^2=-10$, $p_4^2=-10$, $s=200$ and $t=-10$.
The results for both the real part and the imaginary part are compared with  
 those by {\tt XLOOPS-GiNaC}~\cite{xloops} by H.\,S. Do and P.\,H. Khiem. 
The detailed explanation was presented by H.\,S. Do in CPP 2010~\cite{son-cpp2010}, 
showing not only their results and ours, but also results by {\tt LoopTools2.5} with the code {\tt D0C} 
developed by D.\,T. Nhung and L.\,D. Ninh {\it et al.}~\cite{D0C}. All the results are in good agreement.

Through the reported experiences we validated the {\it Direct Computation Method} 
for these loop integrals with real or complex masses of the internal particles.
\subsection{Two-loop box integrals with masses}\label{subsec:2lb}

\begin{figure}[tbp]
\begin{center}
\begin{minipage}[b]{0.49\textwidth}
   \begin{center}
   \includegraphics[width=6.0cm]{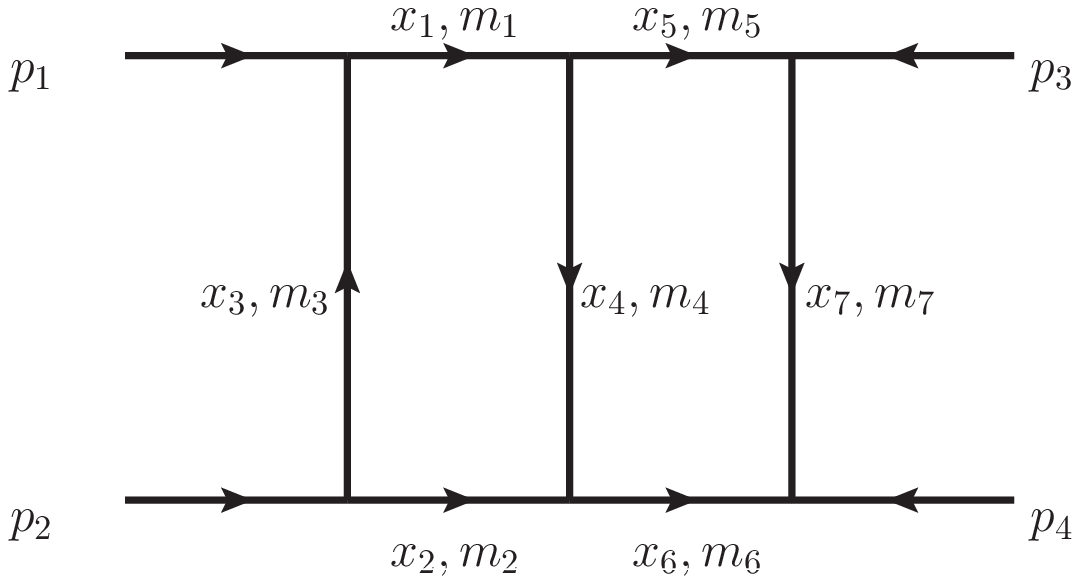}
   \caption{Two-loop planar box diagram}
   \label{fig:2lb-ladder}
   \end{center}
\end{minipage}   
\begin{minipage}[b]{0.49\textwidth}
   \begin{center}
   \includegraphics[width=6.0cm]{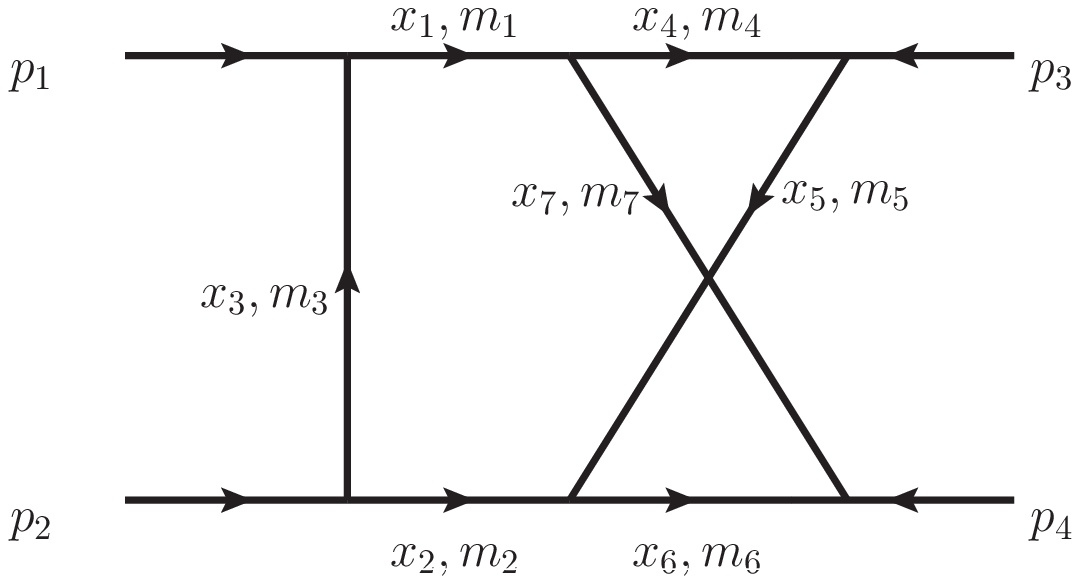}
   \caption{Two-loop non-planar box diagram}
   \label{fig:2lb-cross}
   \end{center}
\end{minipage}   
\end{center}
\vskip-\lastskip
\label{fig:2lb}
\end{figure}

\noindent
Here we discuss the computation of the two-loop box diagram for $L=2$ and $N=7$ in Table~\ref{tab:integrand}.
Corresponding diagrams are shown in Fig.~\ref{fig:2lb-ladder} and in Fig.~\ref{fig:2lb-cross}.
In the following let us consider the scalar loop integral defined as

\begin{equation}\label{eqn:integral-2lb}
I(s,t) = -\int_{0}^{1} dx_{1} ~dx_{2} ~dx_{3} ~dx_{4} ~dx_{5} ~dx_{6} ~dx_{7}
 ~\delta(1-\sum_{\ell=1}^{7}x_{\ell})\frac{C}{(D-i\epsilon C)^{3}}.
\end{equation}

\noindent
$D$ and $C$ in the integrand for the two-loop planar box in Eq.\,(\ref{eqn:integral-2lb}) are given by
\begin{eqnarray*}
{D}&=& {C} \sum_{\ell=1}^{7} x_\ell m^2_\ell \\
            &-&\{s (x_{1}x_{2}(x_{4} + x_{5} + x_{6} + x_{7})+x_{5}x_{6}(x_{1} + x_{2} + x_{3} + x_{4}) + x_{1}x_{4}x_{6} + x_{2}x_{4}x_{5}) \\
           &+& t x_{3}x_{4}x_{7} \\
           &+& p_{1}^2(x_{3}(x_{1}x_{4} + x_{1}x_{5} + x_{1}x_{6} + x_{1}x_{7} + x_{4}x_{5})) \\
           &+& p_{2}^2(x_{3}(x_{2}x_{4} + x_{2}x_{5} + x_{2}x_{6} + x_{2}x_{7} + x_{4}x_{6})) \\
           &+& p_{3}^2(x_{7}(x_{1}x_{4} + x_{1}x_{5} + x_{2}x_{5} + x_{3}x_{5} + x_{4}x_{5})) \\
           &+& p_{4}^2(x_{7}(x_{1}x_{6} + x_{2}x_{4} + x_{2}x_{6} + x_{3}x_{6} + x_{4}x_{6}))\},
\end{eqnarray*}
and
\begin{eqnarray*}
{C}=(x_{1} + x_{2} + x_{3} + x_{4})(x_{4} + x_{5} + x_{6} + x_{7})-x_{4}^2.
\end{eqnarray*}

\noindent
Furthermore, $D$ and $C$ in the integrand for the two-loop non-planar box are given by

\begin{eqnarray*}
{D}&=& {C} \sum_{\ell=1}^{7} x_\ell m^2_\ell \\
            &-&\{s (x_{1}x_{2}x_{4} + x_{1}x_{2}x_{5} + x_{1}x_{2}x_{6} + x_{1}x_{2}x_{7} + x_{1}x_{5}x_{6} + x_{2}x_{4}x_{7} - x_{3}x_{4}x_{6}) \\
            &+& t (x_{3}(- x_{4}x_{6} + x_{5}x_{7}))\\
            &+& p_{1}^2(x_{3}(x_{1}x_{4} + x_{1}x_{5} + x_{1}x_{6} + x_{1}x_{7} + x_{4}x_{6} + x_{4}x_{7}))\\
            &+& p_{2}^2(x_{3}(x_{2}x_{4} + x_{2}x_{5} + x_{2}x_{6} + x_{2}x_{7} + x_{4}x_{6} + x_{5}x_{6}))\\
            &+& p_{3}^2(x_{1}x_{4}x_{5} + x_{1}x_{5}x_{7} + x_{2}x_{4}x_{5} + x_{2}x_{4}x_{6} + x_{3}x_{4}x_{5} + x_{3}x_{4}x_{6} + x_{4}x_{5}x_{6} + x_{4}x_{5}x_{7})\\
            &+& p_{4}^2(x_{1}x_{4}x_{6} + x_{1}x_{6}x_{7} + x_{2}x_{5}x_{7} + x_{2}x_{6}x_{7} + x_{3}x_{4}x_{6} + x_{3}x_{6}x_{7} + x_{4}x_{6}x_{7} + x_{5}x_{6}x_{7}) \},
\end{eqnarray*}
and
\begin{eqnarray*}
{C}=(x_{1} + x_{2} + x_{3} + x_{4} + x_{5})(x_{1} + x_{2} + x_{3} + x_{6} + x_{7}) -(x_{1} + x_{2} + x_{3})^2.
\end{eqnarray*}

\noindent
For both diagrams we have that $s=(p_{1}+p_{2})^2=(p_{3}+p_{4})^2$, ~$t=(p_{1}+p_{3})^2=(p_{2}+p_{4})^2$ and
$p_{1}+p_{2}+p_{3}+p_{4}=0.$

\begin{figure}[ht]
\begin{center}
\includegraphics[width=8.0cm]{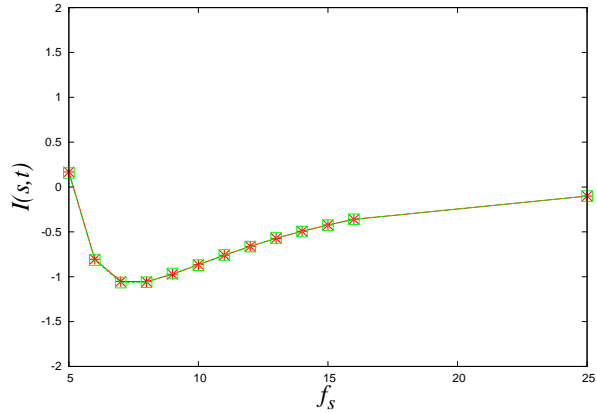}
\end{center}
\caption{Two-loop planar box: Numerical results of real part in units of $10^{-12}$ $\rm GeV^{-6}$ for $5 \leq f_s \leq 25$. Marks (red $\ast$) present results by the {\it Direct Computation Method} and  marks (green $\Box$) are results by the reduction method.}
\label{fig:2lbl-result}
\end{figure}

\begin{figure}[ht]
\begin{center}
\includegraphics[width=8.0cm]{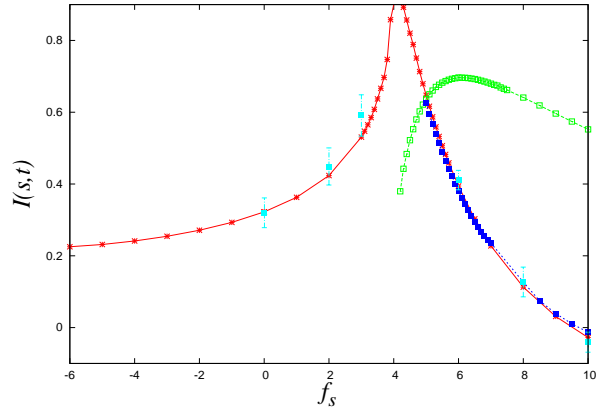}
\end{center}
\caption{Two-loop non-planar box: Numerical results of real and imaginary part in units of $10^{-12}$ $\rm GeV^{-6}$ for $-6 \leq f_s \leq 10$. Marks (red $\ast$) are results for the real part and marks (green $\Box$) for the imaginary part. Marks  (light blue $\Box$) with error-bars present results by the reduction method. Marks (dark blue $\Box$) give numerical results by the dispersion relation for $5 \leq f_s \leq 10$.}
\label{fig:2lbc-result}
\end{figure}

Since the number of dimensions of the integration is 6 and the behavior of the integrand is very complex, we performed
a variable transformation of the Feynman parameters before numerical integration.
With a suitable transformation for each diagram~\cite{jocs}, we completed the computation of the real part of the two-loop planar box integral with $m_{1}=m_{2}=m_{5}=m_{6}=m$ and $m_{3}=m_{4}=m_{7}=M;$ and the real and imaginary part of the non-planar box integral with $m_{1}=m_{2}=m_{4}=m_{6}=m$ and $m_{3}=m_{5}=m_{7}=M$. 
For both $p_1^2=p_2^2=p_3^2=p_4^2=m^2$.
The numerical results for each diagram are plotted as a function of $f_s = s/m^2$ in Fig.~\ref{fig:2lbl-result} and in Fig.~\ref{fig:2lbc-result}, respectively.
For both cases, the kinematical parameters are $t=-100.0^2 {\rm GeV}^2$, $m=50.0 \,{\rm GeV}$ and $M=90.0 \,{\rm GeV}$.


We cross-checked the numerical computation in two ways. First we made a comparison with the reduction method~\cite{iccsa2010}. After reduction we get numerical results using the Monte Carlo integration package {\tt BASES}~\cite{bases}. 
For the two-loop planar box both the results by the {\it Direct Computation Method} and those by the reduction method are plotted in Fig.~\ref{fig:2lbl-result}.
The agreement is very good in the range of interest, $5 \leq f_s \leq 25$. 

For the two-loop non-planar box, both the real part for the whole range and the imaginary part for the range $4 <f_s \leq 10$ are plotted in Fig.~\ref{fig:2lbc-result}.
For the cross-check we attempted the reduction method. However, the behavior of the convergence by Monte Carlo integration is not very good for the intended range.
Therefore six  numerical results by the reduction method are plotted with error-bars in Fig.~\ref{fig:2lbc-result}. Instead of the reduction for the cross-check,  
we performed a self consistency check using the dispersion relation between the real part and the imaginary part,
\begin{equation}\label{eqn:dispersion}
\Re e~I(s) = \frac{1}{\pi}\int\frac{\Im m(I(s'))}{s-s'-i\epsilon}ds'.
\end{equation}
The results for the real part constructed numerically from the imaginary part using Eq.\,(\ref{eqn:dispersion}) are also plotted in Fig.~\ref{fig:2lbc-result} for the range $5 \leq f_s \leq 10$.

\section{Summary}\label{sec:summary}

\noindent
In this paper, we presented a complete numerical approach for multi-loop integrals. From a technical point of view, it is based on a combination of numerical integration and numerical extrapolation and we call the method a 
{\it Direct Computation Method}. 
Some numerical results are given as examples of the technique.
Furthermore, for each numerical result a cross-check has been presented.
We demonstrated the applicability of our approach to loop integrals of the one-loop box diagram with both real and complex masses and two-loop planar and non-planar box diagrams with 
masses. 
~\\
~\\
{\large{\bf{Acknowledgements}}}\\
We thank Prof. T. Kaneko for his valuable suggestions. 
We also thank Dr. H.\,S. Do, Dr. L.\,D. Ninh and Mr. P.\,H. Khiem for their collaboration.
This work is supported in part by the Hayama Center
for Advanced Studies at the Graduate University for Advanced Studies.


\begin{thebibliography}{6}
\small
%
\bibitem{tkachov}
F.\,V. Tkachov, 
Nucl. Instr. Meth. {\bf A 389} (1997) 309.
%
\bibitem{dq1}
E. de Doncker, Y. Shimizu, J. Fujimoto and F. Yuasa, 
{Comput. Phys. Comm.} {\bf 159} (2004) 145.
%
\bibitem{dq2}
E. de Doncker {\it et al.},
Nucl. Instr. Meth. {\bf A 539} (2004) 269.
%
\bibitem{dq3}
E.de Doncker {\it et al.},
{Lecture Notes in Computer Science} {\bf 3514}, (2005) 151-171.
%
\bibitem{dq4}
F. Yuasa {\it et al.}, PoS(ACAT)087.
%
\bibitem{dq5}
E. de Doncker {\it et al.},
talk at LoopFest V, June 19-21, 2006.
%
\bibitem{acat08}
F. Yuasa {\it et al.}, PoS(ACAT08)122.
%
\bibitem{1loop1}
Y. Oyanagi {\it et al.},
in {\it Perspectives of Particle Physics},World Scientific 1989 ISBN 9971-50-589-4, p.369.
%
\bibitem{supplement}
J. Fujimoto {\it et al.},
Prog. Theor. Phys. Suppl. {\bf No.100} (1990) 1.
%
\bibitem{1loop2}
J. Fujimoto {\it et al.},
in proceedings of {\it Computing in High Energy Physics '91}, Universal Academy Press, Inc., Tokyo, Japan 1991, p.407.
%
\bibitem{1loop3}
J. Fujimoto, Y. Shimizu,  K. Kato and Y. Oyanagi,
Prog. Theor. Phys. Suppl. {\bf Vol.87} No. 5 (1992) 1233.
%
\bibitem{kreimer}
D. Kreimer, Phys. Lett. {\bf B273} (1991)277.
%
\bibitem{2loopself}
J. Fujimoto, {\it et al.},
in {\it New Computing Techniques in Physics Research II}, World Scientific, 1992, p.625.
%
\bibitem{2loopv1}
J. Fujimoto {\it et al.},
in {\it Proc. of VII th Workshop on High Energy Physics and Quantum Filed Theory},
Sochi, Russia, Oct 7-14, 1992.
%
\bibitem{2loopv2}
J. Fujimoto {\it et al.},
Int. J. of Mod. Phys. C, {\bf Vol.6} No. 4 (1995)525.
%
\bibitem{tarasov}
O.\,V. Tarasov,
in {\it New Computing Techniques in Physics Research IV}, World Scientific, 1995, p.161.
%
\bibitem{bauberger}
S. Bauberger {\it et al.}, Nucl. Phys. {\bf B445} (1995) 25.
%
\bibitem{passarino-self}
G. Passarino {\it et al.}, Nucl. Phys. {\bf B629} (2002) 97.
%
\bibitem{passarino-2lv}
A. Ferroglia {\it et al.}, Nucl. Phys. {\bf B680} (2004) 190-270.
%
\bibitem{nci}
Y. Kurihara and T. Kaneko,
Comput. Phys. Comm. {\bf 174} (2006) 530.
%
\bibitem{ueda-acat08}
T. Ueda {\it et al.}, PoS(ACAT08)120 and private communication. 
%
\bibitem{quadpack}
R. Piessens, {\it et al.},
\emph{QUADPACK, A Subroutine Package for Automatic Integration},
Springer Series in Computational Mathematics. Springer-Verlag, 1983.
%
%
\bibitem{iterate}
S. Li, E. de Doncker and K. Kaugars,
{Lecture Notes in Computer Science} {\bf 3514}, (2005) 123-130.
%
\bibitem{wynn}
P. Wynn,
{Mathematical Tables and Other Aids to Computation} {\bf 10} No. {\bf 54}(1956) 91, and 
SIAM J. Numer. Anal. {\bf 3} (1966) 91.
%
\bibitem{FF}
G.\,J. van Oldenborgh and J.\,A.\,M. Vermaseren,
{Z. Phys.} {\bf C46}, 425 (1990).\\
G.\,J. van Oldenborgh, 
Comput. Phys. Comm. {\bf 66} (1991) 1-15.
%
\bibitem{looptools}
T. Hahn,
{Nucl. Phys. Proc. Suppl.} {\bf 89} 231-236, (2000).
%
\bibitem{bbh}
F. Boudjema and L.\,D. Ninh, 
Phys. Rev. {\bf D 78}, 093005 (2008).
%
\bibitem{xloops}
C. Bauer,
in proceedings of {\it CPP 2001}, 
27-30 Nov. 2001, 
({\bf KEK Proceedings 2002-11}, pp.179-185), \\ 
H.S. Do,
Ph.D thesis at the Physics Department, Johannes Gutenberg University Mainz, \\
C. Bauer and H.\,S. Do,
Comput. Phys. Comm. {\bf 144} (2002) 154-168.
%
\bibitem{D0C}
D.\,T. Nhung and L.\,D.Ninh, Comput. Phys. Comm. {\bf 180} (2009)2258.
%
%
%
%
%
\bibitem{son-cpp2010}
H.\,S. Do, P.\,H. Khiem and F. Yuasa, the proceedings of {\it CPP2010}, to appear in PoS.
%
\bibitem{iccsa2010}
E. de Doncker {\it et al.}, 
{Lecture Notes in Computer Science} {\bf 6017}, (2010) 139-154.
%
\bibitem{bases}
S. Kawabata,
Comput. Phys. Comm. {\bf 88} (1995) 309-326.
%
\bibitem{jocs}
E. de Doncker {\it et al.},
submitted to {Journal of Computational Science}.
%
\end{thebibliography}
\end{document}